\documentclass[10pt,letterpaper,twoside]{cpc-hepnp}
\usepackage{amsmath}
\usepackage{amssymb}
\usepackage{graphicx}

\makeatletter

\pdfpageheight\paperheight
\pdfpagewidth\paperwidth

\usepackage{multicol}\usepackage{bm}\usepackage{mathrsfs}\usepackage{bbm}\usepackage{amscd}\usepackage{lastpage}\usepackage{CJK}

\makeatother

\begin{document}
\begin{CJK*}{GBK}{song}

\fancyhead[c]{\small To be submitted to Chinese Physics C} \fancyfoot\thepage

\footnotetext[0]{Received XX XXX 2016}

\title{{\normalsize{}Compensation for Booster Leakage Field in the Duke
Storage Ring }\thanks{Supported by National Natural Science Foundation of China (11175180,
11475167) and US DOE (DE-FG02-97ER41033) }{\normalsize{} }}

\author{{Wei Li$^{1,2;1)}$\email{wei.li3@duke.edu}
    \quad{}Hao Hao$^{2,2)}$\email{hhao@fel.duke.edu}
    \quad{}Stepan F. Mikhailov$^{2}$
    \quad{}Victor Popov$^{2}$}\\
    Wei-Min Li$^{1}$
    \quad{}Ying K. Wu$^{2}$}

\maketitle
\address{%
$^1$ National Synchrotron Radiation Laboratory, University of Science and Technology
of China, Hefei, 230029, China\\
$^2$ Triangle University Nuclear Laboratory/Physics Department, Duke University, Durham, 27705, USA\\
}
\begin{abstract}
{\normalsize{}The High Intensity Gamma-ray Source (HIGS) at Duke University
is an accelerator-driven Compton gamma-ray source, providing high
flux gamma-ray beam from 1 MeV to 100 MeV for photo-nuclear physics
research. The HIGS facility operates three accelerators, a linac pre-injector
(0.16 GeV), a booster injector (0.16--1.2 GeV), and an electron storage
ring (0.24--1.2 GeV). Because of proximity of the booster injector
to the storage ring, the magnetic field of the booster dipoles close
to the ring can significantly alter the closed orbit in the storage
ring being operated in the low energy region. This type of orbit distortion
can be a problem for certain precision experiments which demand a
high degree of the energy consistency of the gamma-ray beam. This
energy consistency can be achieved by maintaining consistent aiming
of the gamma-ray beam, therefore, a steady electron beam orbit and
angle at the Compton collision point. To overcome the booster leakage
field problem, we have developed an orbit compensation scheme. This
scheme is developed using two fast orbit correctors and implemented
as a feedforward which is operated transparently together with the
slow orbit feedback system. In this paper, we will describe the development
of this leakage field compensation scheme, and report the measurement
results which have demonstrated the effectiveness of the scheme.}{\normalsize \par}
\end{abstract}
\begin{keyword} field compensation, feedforward, storage ring, beam
orbit\end{keyword}

$\text{\,\,}$PACS: 29.20.db

\footnotetext[0]{\hspace*{-3mm}\raisebox{0.3ex}{${\scriptstyle \copyright}$}2013
Chinese Physical Society and the Institute of High Energy Physics
of the Chinese Academy of Sciences and the Institute of Modern Physics
of the Chinese Academy of Sciences and IOP Publishing Ltd}

\begin{multicols}{2}

\section{{\normalsize{}Introduction}}

The High Intensity Gamma-ray Source (HIGS) at Duke University is an
accelerator-driven Compton gamma-ray source for photo-nuclear research.
It provides a very high intensity gamma-ray beam with switchable polarization
in the energy range from 1 MeV to 100 MeV. The HIGS facility consists
of three accelerators: a linac pre-injector (0.16 GeV), a booster
injector (0.16--1.2 GeV), and an electron storage ring (0.24--1.2
GeV) \cite{Stepan2007,Stepan2009}, as shown in Fig. \ref{fig:1-1-Layout-of-DukeStorageRing}.
The 34-meter-long south straight section hosts the Duke Free-Electron
Lasers (FELs) with several undulator configurations, producing an
FEL beam with wavelength ranging from 190 nm to 2 $\mu$m \cite{YKWu2006}.
The FEL beam is trapped in a 53.73 m long low-loss laser cavity to
be synchronized with the circulating electron beam (2.79 MHz). Operating
in the two-bunch mode with two electron bunches separated by a half
of the storage ring circumference from each other, a high flux gamma-ray
beam is produced in the middle of the south straight section by colliding
one electron bunch head-on with the high power FEL beam produced by
the other electron bunch \cite{InvCompScat-1,InvCompScat-2}. After
collimated by the downstream collimator, the gamma-ray beam becomes
quasi-monochromatic and is delivered to the target room.

However, due to the proximity of the booster injector to the storage
ring, the magnetic field of the booster dipoles close to the storage
ring can significantly alter the closed orbit in the storage ring
when being operated in the low energy region. The resulted beam orbit
distortion can be a significant problem for low energy gamma-ray production,
which requires consistent aiming of the gamma-ray beam, therefore,
a steady electron beam orbit in the collision area. This kind of disturbance
can be a problem for similar accelerator facilities in which the booster
injector shares the same tunnel with the storage ring \cite{WJoho2006,TPS2015}.
This kind of problem is also very important for the next generation
light source, e.g. diffraction limited storage ring (DLSR) \cite{Erikson2014},
since electron beams in the DLSR can be sensitive to the integrated
magnetic field variation on the level of $10^{-6}\text{ T}\cdot$m.
Therefore, it is important to develop shielding or compensation schemes
for the leakage magnetic field from a nearby ramping booster \cite{WJoho2006,RKiller1997}.

At the HIGS facility, during routine operation the electron beam is
first injected into the booster at 169 MeV, the booster is then ramped
to bring the electron beam to the extraction energy (any where between
240 MeV and 1.2 GeV), the electron beam is extracted into the storage
ring, and finally the booster is ramped up to the maximum energy of
1.2 GeV and then down to 169 MeV to finish one injection cycle. In
this process, it is observed that the beam orbit in the storage ring
is significantly affected by the booster leakage field, as shown in
Fig. \ref{fig:1-2-dynamic.orbit.during.injection.cycle}. For gamma-ray
operation above about 15 MeV, Compton scattered electrons lose too
much energy so that they cannot be retained in the storage ring \cite{HIGGS.High.Loss_PAC2009}.
In this electron loss mode, continuous electron beam injection into
the storage ring is needed to compensate for a high rate of electron
loss. In this mode, the amount of time during which the collision
orbit is altered due to the periodic booster ramping becomes a significant
portion of the gamma-ray production. The adverse impact of the booster
leakage field is most significant for the Compton gamma-ray production
using a low energy electron beam below 600 MeV.

\noindent \end{multicols} \ruleup

\noindent \begin{center}
\includegraphics[width=17cm]{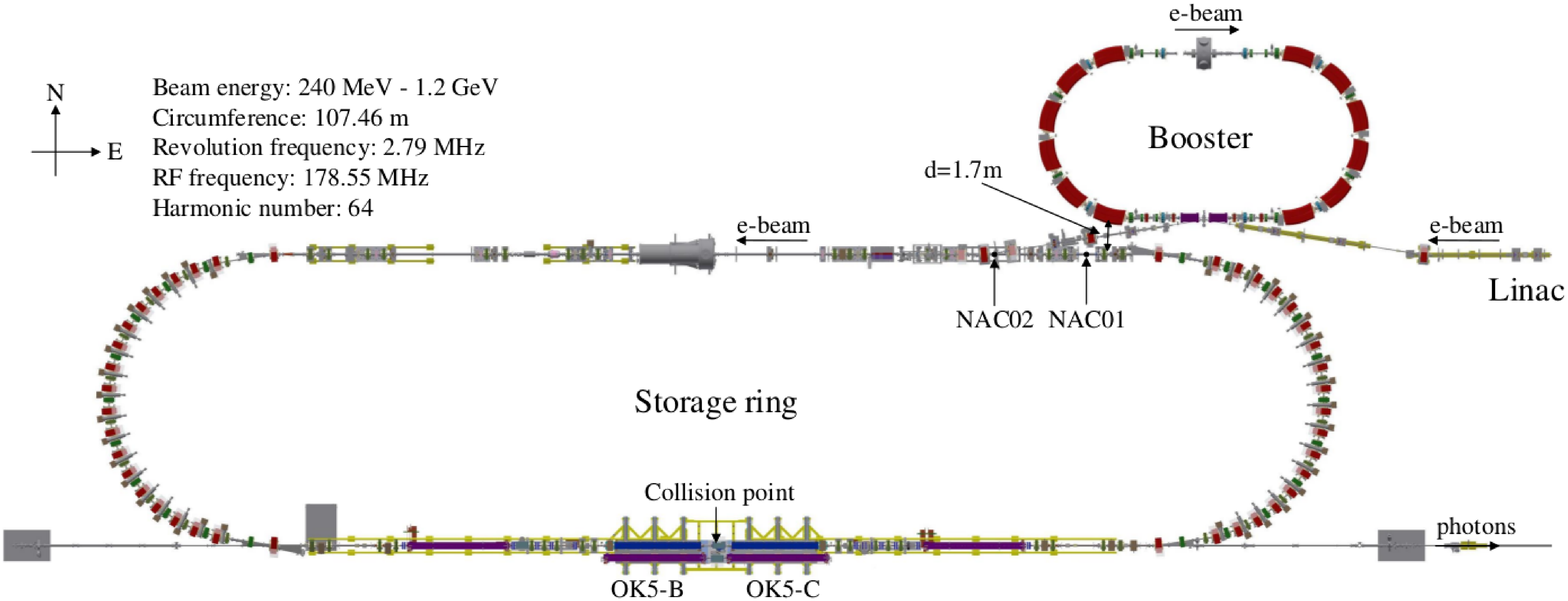}
\figcaption{
The layout of Duke storage ring and its booster injector. The closest
booster dipole magnet is about 1.7m away from the nearby storage ring
beam line. \label{fig:1-1-Layout-of-DukeStorageRing} }
\par\end{center}

\noindent \ruledown \begin{multicols}{2}

\noindent \begin{center}
\includegraphics[width=8cm]{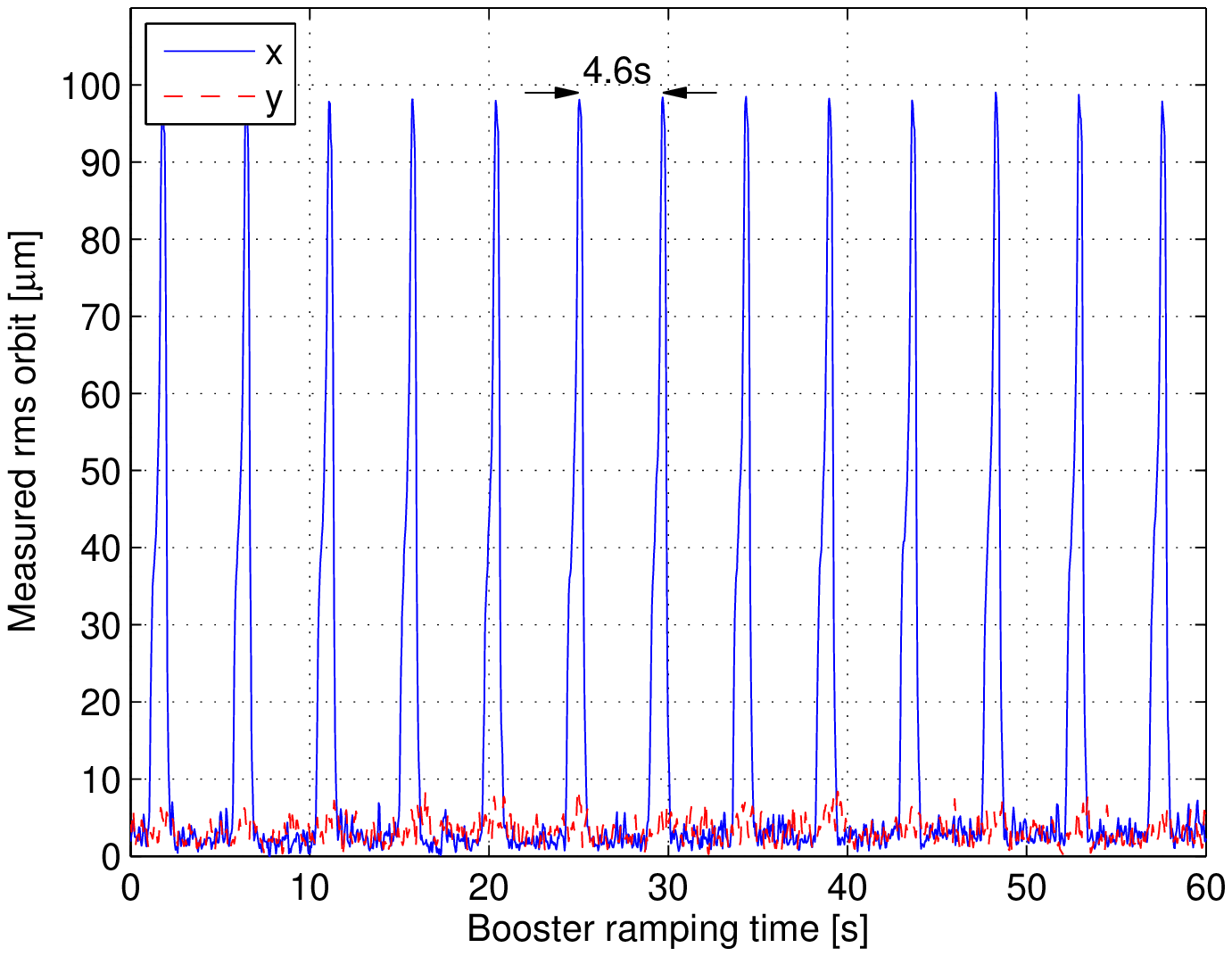}\figcaption{
(color online) The measured rms beam orbit deviations in the storage
ring as a function of time while continuously ramping the booster
with a repetition rate 4.6 s. The storage ring is operated at $E_{SR}=426$
MeV. \label{fig:1-2-dynamic.orbit.during.injection.cycle} }
\par\end{center}

To obtain a consistent collision orbit for the gamma-ray operation,
an orbit compensation scheme for the booster leakage field is developed.
This scheme is devised using a beam based technique with two air coil
correctors, and implemented as a feedforward system which is operated
transparently together with the slow orbit feedback system. In this
paper, we will present the development of the orbit compensation scheme
for the booster leakage field, and report the measurement results
of this compensation in two different modes: (1) the static mode of
operation while stepping the booster energies, and (2) the dynamic
mode of operation with continuous booster ramping.

\section{{\normalsize{}Field leakage compensation scheme}}

The closed orbit in a circular accelerator is determined by the magnetic
field (i.e. in magnets) and electric field (e.g. in the RF cavity)
around the accelerator, which can be distorted by the field errors.
To control the orbit distortions caused by slow and distributed magnetic
field errors, a slow orbit feedback is commonly employed, in which
the corrector strengths are calculated based on the beam orbit deviations
measured using beam position monitors (BPMs) and a pre-measured response
matrix.

At the Duke storage ring (DSR), the slow orbit feedback system is
composed of 55 horizontal correctors, 24 vertical correctors and 32
BPMs \cite{BPM-Corrector}. It is operated at 1 Hz, which is fast
enough to correct the orbit distortions caused by slow varying errors,
e.g. those associated with temperature changes and slow power supply
drifts. For the booster injector, the full energy ramp up and down
cycle takes about 1.2 s to complete, hence it is not feasible to control
the orbit perturbation caused by the booster leakage field using the
existing slow orbit feedback system. The fast orbit feedback operating
at a few hundred or thousand Hz may be feasible for the booster orbit
correction \cite{FOFB1999}, but it requires a significantly investment
on the electronics and vacuum system. Hence, an alternative economic
compensation scheme is needed. Since the booster is located to the
northeast of the storage ring, the leakage field is also confined
to this area. To minimize the orbit perturbation outside of this area,
some local correctors can be employed. Moreover, as the strength of
the leakage field is repeatable and can be pre-determined, a feedforward
correction scheme can be used. In this scheme, the corrector strengths
can be determined in advance as a function of the leakage field strength
or the booster energy, and correctors can be synchronized with the
booster energy ramping.

In the HIGS operation, a nearly monochromatic gamma-ray beam is produced
by sending the beam through a small collimator. The aiming of the
gamma-ray beam is determined by the position and direction of the
relativistic electron beam at the collision point. Therefore, to maintain
the energy resolution of the gamma-ray beam, the change of the electron
beam orbit (both displacement and angle) at the collision point must
be kept small. For some electron beam displacement $\Delta x_{c}$
and angular deviation $\Delta\theta_{c}$ at the collision point,
the offset of the gamma-ray beam center $\Delta x_{coll}$ at the
location of the collimator is given by

\begin{equation}
\Delta x_{coll}=\Delta x_{c}+\Delta\theta_{c}\cdot L,\label{eq:1-1-distortion.on.collimator}
\end{equation}
where $L=53$ m is the distance between the collision point and the
collimator. If we require $\Delta x_{coll}$ to be less than 10\%
of the smallest collimator diameter (6 mm), the maximumly allowed
electron beam displacement is $\Delta x_{coll,max}=0.6$ mm, and the
maximumly allowed angular variation is $\Delta\theta_{coll,max}=11\mu\text{rad}$.
Typically, the electron orbit displacement is small, the main problem
is the change of the electron beam angle at the collision point, as
its effect is significantly amplified by the large distance between
the collision point and the location of the collimator ($L=53$ m).

The perturbed orbit in the storage ring is measured with the booster
injector parked at 1.2 GeV (the maximum energy), as shown in Fig.
\ref{fig:2-1-Perturbed.eletron.beam.orbit}. It is observed that the
perturbation is mostly in the horizontal direction with the maximum
offset of 179 $\mu$m around the storage ring. In this case, the orbit
displacement is about 115 $\mu$m and angular variation is about 14
$\mu$rad at the collision point. In vertical direction, the orbit
perturbation is one order of magnitude smaller, which is too small
to be a problem. Thus, the compensation scheme will focus on reducing
the horizontal orbit distortion. Scaling this measurement to the lowest
energy operation of the storage ring (240 MeV), to achieve the tight
tolerance set for the gamma-ray beam collimation established previously,
our goal is to reduce the electron beam orbit displacement and angle
at the collision point by a factor of about 2.5.

\noindent \begin{center}
\includegraphics[width=8cm]{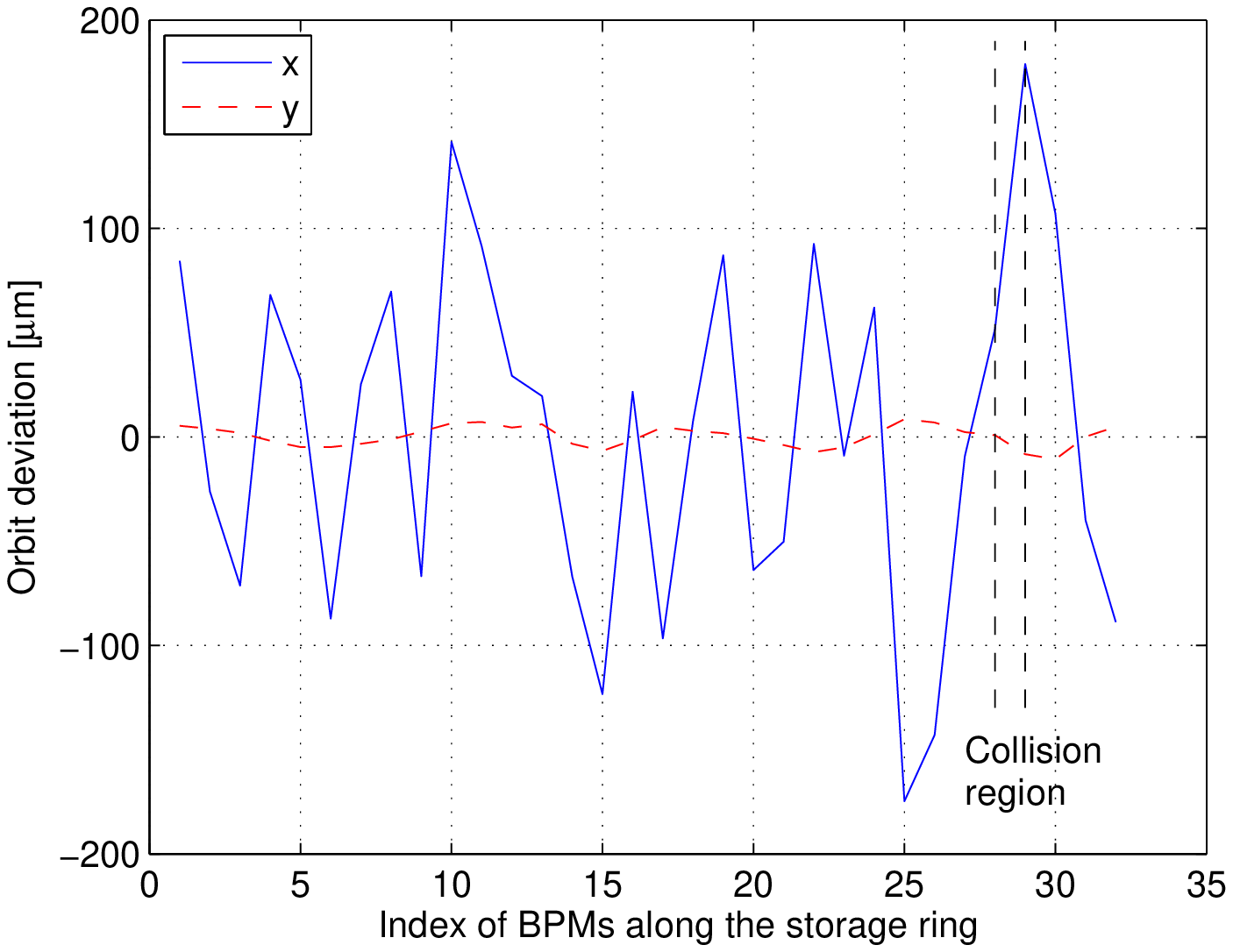}\figcaption{
(color online) Perturbed electron beam orbit along the storage ring.
Measured with the storage ring energy $E_{SR}=426\text{ MeV}$ and
the booster energy $E{}_{BST}=1.2\text{ GeV}$.\label{fig:2-1-Perturbed.eletron.beam.orbit}}
\par\end{center}

\noindent \begin{center}
\includegraphics[width=8cm]{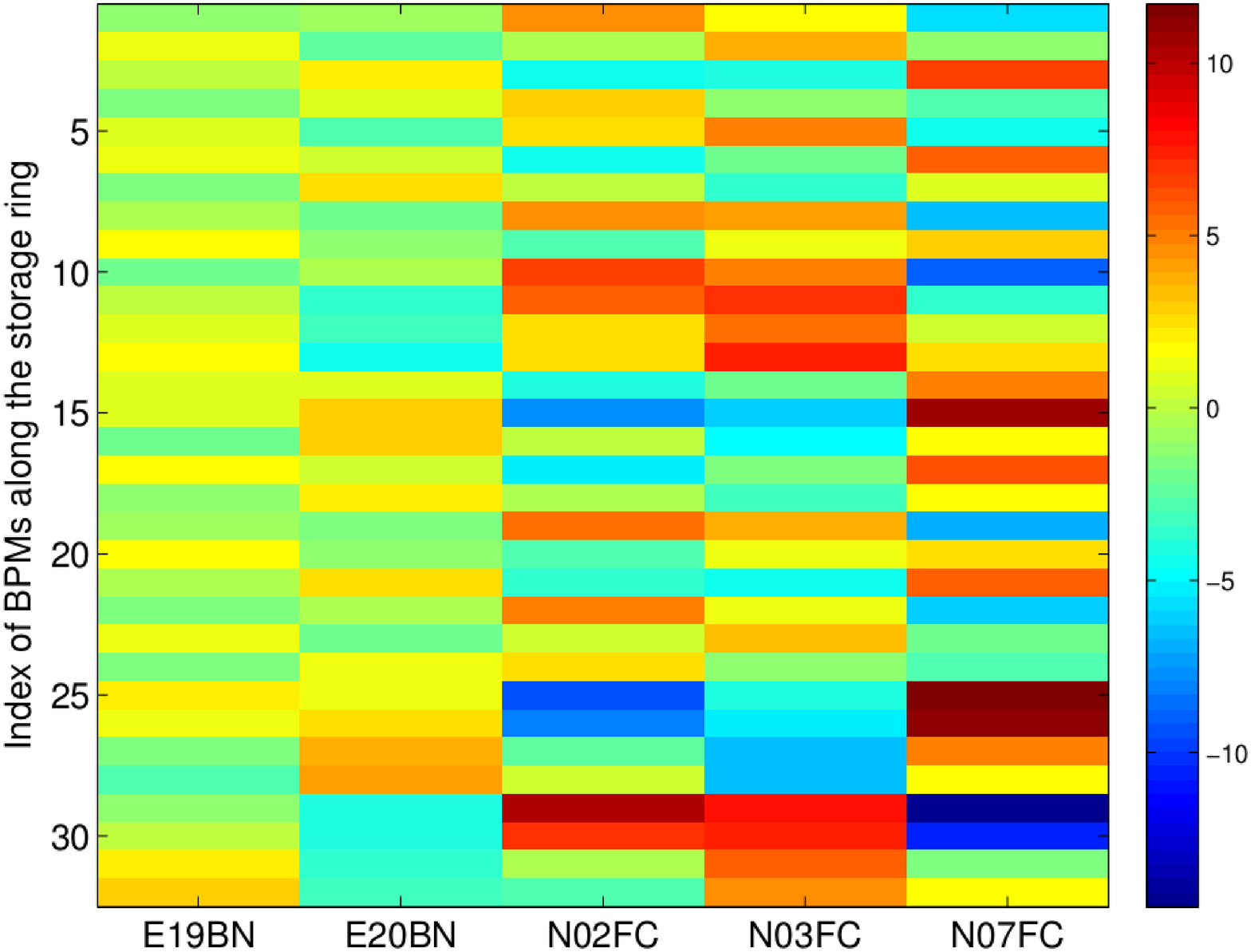}\figcaption{
(color online) Response matrix of a few selected correctors in the
northeast section, measured with $E_{SR}=426\text{ MeV}$, $E_{BST}=500\text{ MeV}$.\label{fig:2-2-Resp.Mat}}
\par\end{center}

With the booster ramped to the maximum energy of 1.2 GeV (Fig. 3),
the effective kicking angle, assuming the orbit disturbance is localized,
can be estimated to be about $\theta_{0}\thickapprox20\,\mu$rad ($E_{SR}=426\text{ MeV}$),
or an integrated field about $2.9\times10^{-5}\text{ T}\cdot\text{m}$
\cite{SYLee}. This correction is small, comparable to the strength
of a typical corrector in the orbit feedback system. It is also noticed
that the horizontal phase advance in the northeast area of the storage
ring is $\Delta\mu_{x}\thickapprox0.6\pi$, therefore only a few correctors
in this area are needed to minimize the orbit perturbation both inside
and outside the area of the booster leakage field.

Five correctors in the northeast area of the storage ring are chosen
as the candidates to compensate the localized booster leakage field.
A beam based technique is used to determine the most effective combination
of correctors. The response matrix for these correctors as shown in
Fig. \ref{fig:2-2-Resp.Mat}. Using the singular value decomposition
(SVD) algorithm, the most effective eigenmode is found to be dominated
by the N07FC and N02FC correctors. It is also known that these two
correctors are separated by a betatron phase difference of $1.1\pi$,
which means they are almost equivalent but with opposite kicking angles.
As the N07FC corrector is located further from the error source area,
the N02FC corrector is clearly the better local corrector. In addition,
a second corrector N03FC which has a $0.45\pi$ phase difference from
N02FC is selected as the secondary corrector to reduce the residual
orbit errors that N02FC is not capable of correcting.

\noindent \begin{center}
\includegraphics[width=8cm]{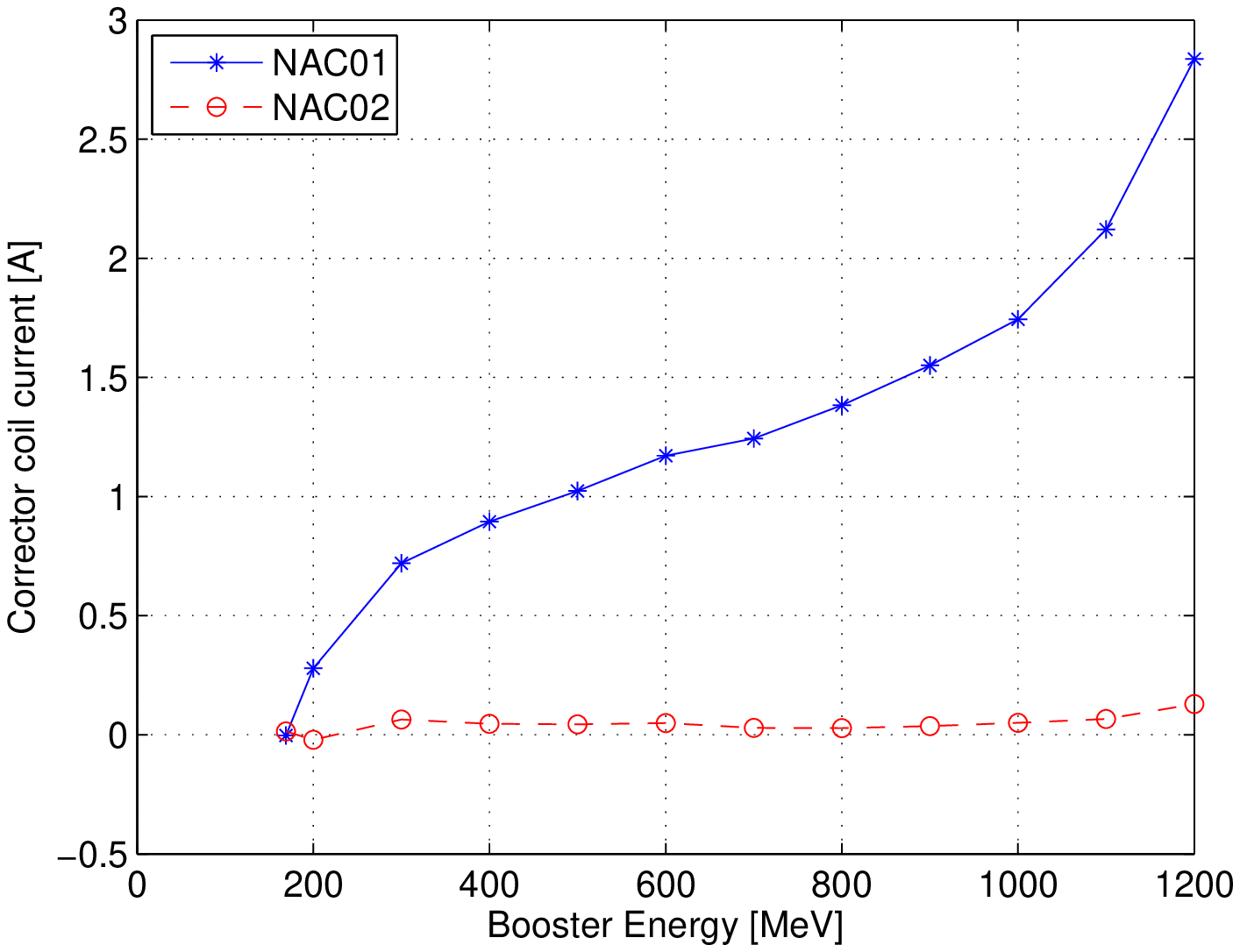}\figcaption{
(color online) Strengths of NAC01 and NAC02 measured with the booster
injector parked at a set of energies between 169 and 1200 MeV. $E{}_{SR}=426\text{ MeV}$.\label{fig:2-3-Strength.of.correctors}
}
\par\end{center}

Since a typical booster energy ramping takes only 1.2 s, fast power
supplies for the correctors are needed. A couple of ramping power
supplies which are properly synchronized with the booster energy ramping
are used to drive the N02FC and N03FC correctors. In the measurement
it was found that the magnetic field produced by these correctors
has a time delay due to the induced eddy current in the soft-iron
yoke. To overcome this problem, two dedicated air coil correctors
NAC01 and NAC02 were developed and installed next to the N02FC and
N03FC correctors, respectively, as shown in Fig. \ref{fig:1-1-Layout-of-DukeStorageRing}.

The strengths of the new correctors are determined with the booster
injector parked at different energies (static mode), as shown in Fig.
\ref{fig:2-3-Strength.of.correctors}. It is clear that the strength
of NAC01 is much larger than that of NAC02, which indicates that the
booster leakage field is mostly local, therefore its main effect can
be corrected using one local corrector NAC01. The maximum integrated
field produced by NAC01 is estimated to be about $2.4\times10^{-5}$
T$\cdot$m, which is close to previously estimated value. It is observed
that the strength of NAC01 is almost linear in the booster energy
range from 300 MeV to 900 MeV, but in the low and high energy regions
a nonlinear dependency on the booster energy shows up, which is likely
resulted from the changed distribution of the leakage field. The compensation
scheme is implemented in the EPICS based control system as a feedforward
control, where the corrector strengths track the booster energy ramping.

\section{{\normalsize{}Compensation results}}

To demonstrate the effectiveness of the compensation scheme, the uncompensated
and compensated electron beam orbits in the storage ring are measured
by the BPMs with the booster operated in both static and dynamic modes.
Its usefulness is further confirmed using the direct measurements
of the aiming stability of the undulator radiation.

\subsection*{{\normalsize{}Static mode}}

\noindent \begin{center}
\includegraphics[width=8cm]{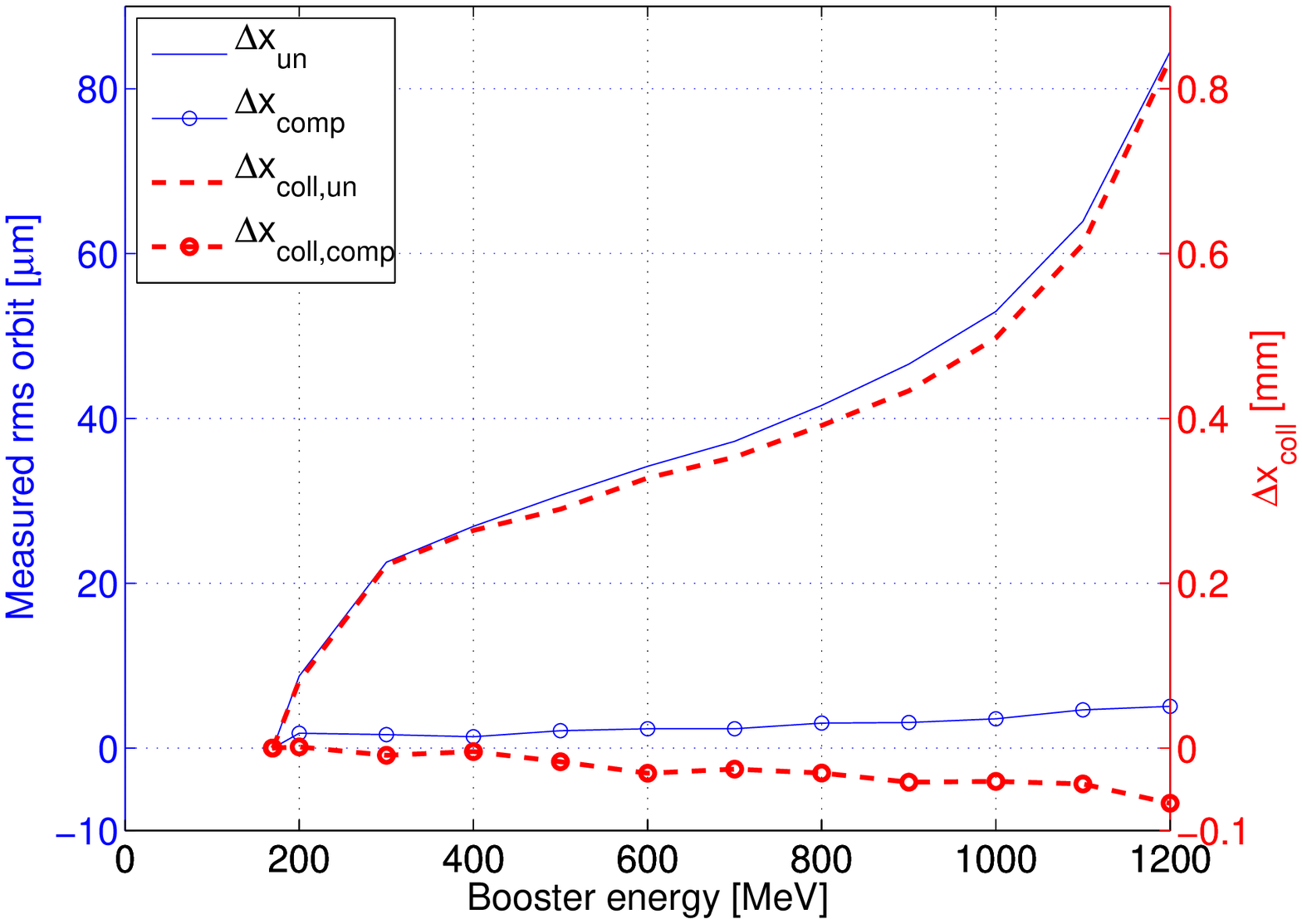}\figcaption{(color
online) The rms beam orbit change in the storage ring and the estimated
gamma-ray center shift at the collimator as a function of the booster
energy with the booster being operated in the static mode. $\Delta x_{un}$
is the rms beam orbit measured without compensation and $\Delta x_{comp}$
is the value with compensation. $\Delta x_{coll,un}$ is the change
of the gamma-ray beam center at the collimator without compensation,
$\Delta x_{coll,comp}$ is the value with compensation. $E{}_{SR}=426$
MeV.\label{fig:3-1-Rmsorb_BefAft_DC}}
\par\end{center}

As shown in Fig. \ref{fig:3-1-Rmsorb_BefAft_DC}, the beam orbit in
the storage ring is measured using all BPMs around the storage ring
with the booster operated in the static modes. The beam orbit offset
$\Delta x{}_{c}$ and angle $\Delta\theta_{c}$ at the collision point
can be calculated using the two BPMs located at the either end of
OK5-B and OK5-C undulator, respectively. Then the variations of the
gamma-ray beam center at the collimator can be obtained using Eq.
\ref{eq:1-1-distortion.on.collimator}. It is observed that the trend
of the storage ring rms orbit variation with the booster energy is
very similar to the variation of the gamma-ray beam center at the
collimator, which indicates a strong correlation between them. The
measurement results show that in the static mode the compensation
scheme reduces the maximum rms orbit variation by a factor of about
15, the same goes to the gamma-ray beam center variation, which is
reduced from 840 $\mu$m to about --67 $\mu$m, well below the stability
requirement for the HIGS operation (see in Section 2).

\subsection*{{\normalsize{}Dynamic mode}}

\noindent \begin{center}
\includegraphics[width=8cm]{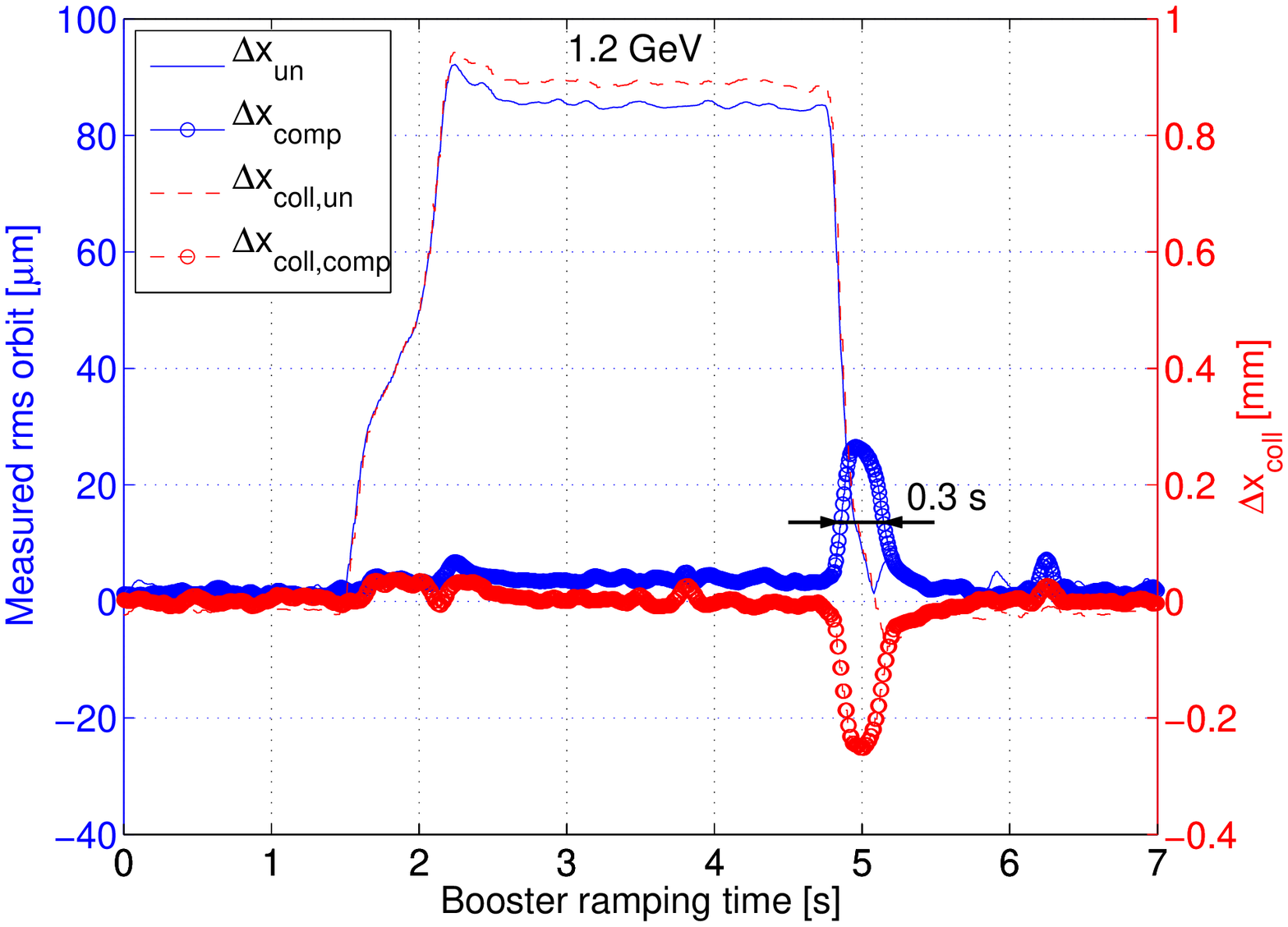}\figcaption{(color
online) The measured rms orbit distortions and estimated gamma-ray
center variations at the collimator during a full booster ramp cycle.
During the cycle, the booster is parked at 1.2 GeV for about $2.5\,s$
(from $t=2.2$ s to $4.7$ s). $E{}_{SR}=426$ MeV.\label{fig:3-2-Rmsorb_BefAft_AC}}
\par\end{center}

At the Duke storage ring, while the slow orbit feedback system uses
the highly filtered low noise orbit data at a few Hz, the BPM acquisition
system is capable of providing data at a higher rate of 30 Hz. The
30 Hz orbit data are fast enough to visualize the storage ring orbit
changes during the 1.2 s booster energy ramp. As shown in Fig. \ref{fig:3-2-Rmsorb_BefAft_AC},
the beam orbit in the storage ring is measured for a complete booster
ramping cycle. In this measurement, the booster starts its ramp from
169 MeV at about $t=1.4$ s and reaches the highest energy of 1.2
GeV at $t=2.2$ s. After parking at 1.2 GeV for 2.5 s, it ramps down
to 169 MeV in about 0.4 s. It is clear that the storage ring orbit
stability is significantly improved along the booster ramp-up curve.
The same is seen in the corresponding shift of the gamma-ray beam
center. In the booster ramp-down process, a bump, with its peak about
30\% of the uncompensated value, shows up in both curves. Even with
this residual orbit bump during the booster energy ramp down, its
amplitude is already small enough to satisfy the established goal
for the compensation scheme (see Section 2). The duration of the residual
bump is also much shorter than the time for the booster extraction
(1 s to 2.5 s). The overall time-integrated orbit bump at the collision
point is reduced by a factor of about 10 to 16 using this realtime feedforward
field compensation, and the time-integrated gamma-ray perturbation
on the collimator is reduced by a factor of about 14 to 25.

\noindent \begin{center}
\includegraphics[width=8cm]{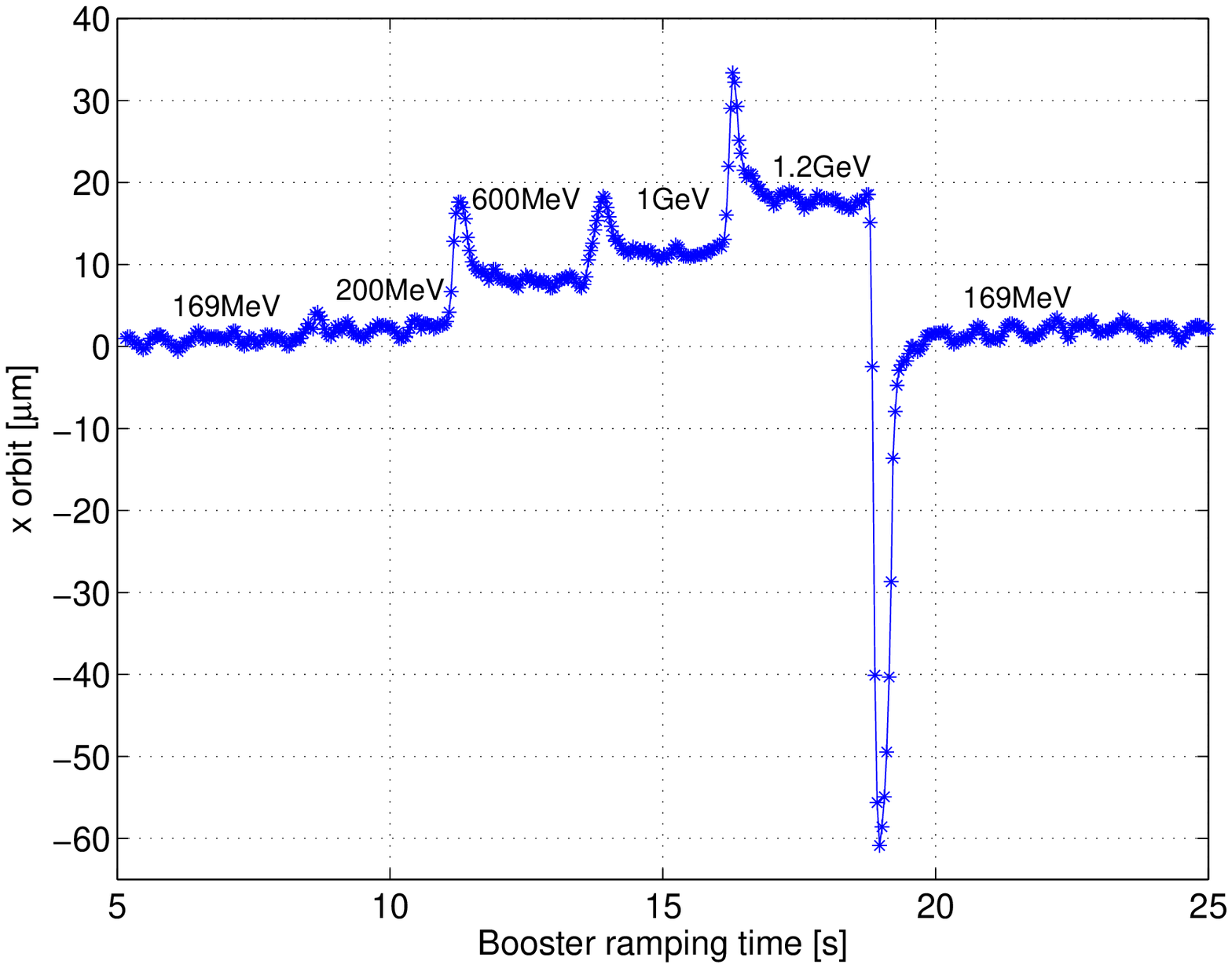}\figcaption{(color
online) Electron beam positions measured using the BPM at S07 sector
with booster ramping. The beam orbit is measured with the booster
ramped up in four discrete energy steps and parked at each energy
for 2.5 s, and then ramped down to complete a cycle. As a reference,
the uncompensated orbit at this location was about $+179$ $\mu$m
with the booster energy parked at 1.2 GeV. $E{}_{SR}=426\text{ MeV}$.\label{fig:3-3-Detailed.view.of.compensated.AC.orbit.standard.deviation}}
\par\end{center}

The storage ring beam orbits were also measured with the booster energy
ramped up from 169 MeV to 1.2 GeV in four discrete steps while parking
at intermediate energies of 200 MeV, 600 MeV, 1 GeV, 1.2 GeV for 2.5
second each, and then ramped down to finish one cycle. The beam positions
measured using the BPM at S07 sector (S07BPM) are shown in Fig. \ref{fig:3-3-Detailed.view.of.compensated.AC.orbit.standard.deviation}.
It is observed that there is a small orbit peak during energy ramp-up
and a small orbit shift after each ramp, both are the results of small
under-correction. It is also observed that the beam orbit bump is
reversed along the ramp-down process, indicating orbit over-correction.
These orbit peaks, either as under-correction during energy ramp-up,
or over-correction during energy ramp-down, are the results of the
delay in the field compensation system. While many factors can contribute
to such delay, including the finite time response of the control system
and corrector power supplies, the most likely factor is the eddy current
effect in the thick storage ring vacuum chamber (3 mm, stainless steel).

\subsection*{{\normalsize{}Direct measurement of undulator radiation aiming}}

Beam orbit stability in the south straight section is further verified
by directly measuring the stability of the FEL undulator radiation
aiming. The OK5-B and OK5-C undulators located adjacent to the Compton
collision point are turned on and set to a fixed current to produce
synchrotron radiation. The undulator photon beam (center wavelength
$\lambda_{cen}=550$ nm, $E_{SR}=426\text{ MeV}$) is measured using
a beam profile monitor located about 30 m downstream from the Compton
collision point. The CCD camera in the monitor is capable of taking
about 14 frames per second in the burst mode. Due to the long distance
between the radiation source and the camera, the variation of the
undulator beam image center at the camera is mainly caused by the
change of the electron beam angle in the undulators. In this measurement,
a bandpass optical filter is used to reduce the background radiation.

\includegraphics[width=8cm]{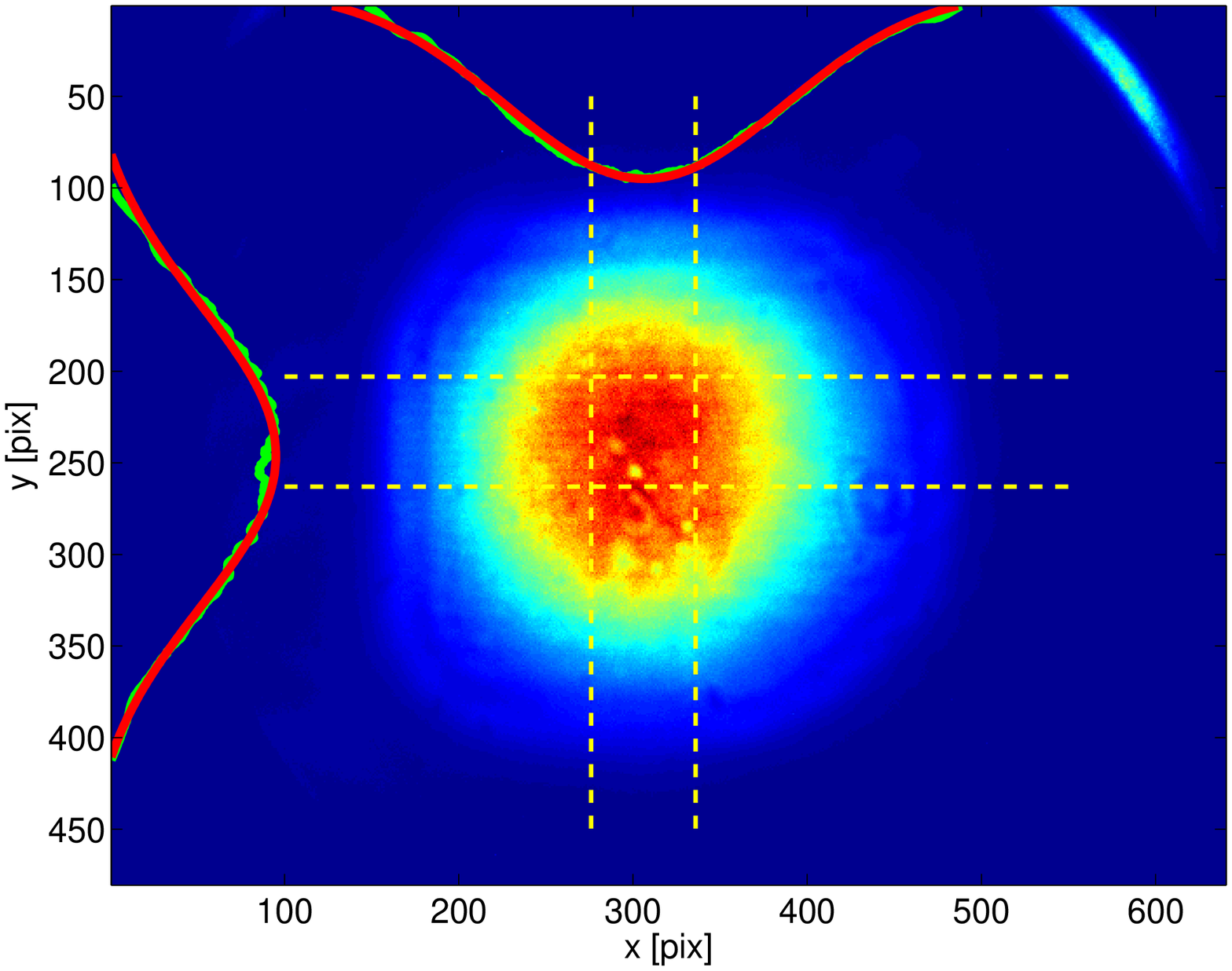}\figcaption{(color
online) Measured FEL undulator radiation beam image with center wavelength
$\lambda_{cen}=550\text{ nm}$. The beam image centers are fitted
with Gaussian functions using the intensity distribution within the
selected areas. The yellow dash-lines: the boundaries of the selected
areas; green dot-lines: the intensity distribution of the data in
the selected areas; red lines: the fitted curve of the intensity distribution.
$E_{SR}=426\text{ MeV}$. \label{fig:3-6-FitCenter}}

The image data are processed first by subtracting the background noise,
and the image center is found by fitting the intensity distribution
of a selected area in the horizontal or vertical direction using a
Gaussian function, as shown in Fig. \ref{fig:3-6-FitCenter}. Hence,
the beam images can be recorded with the booster operated in the dynamic
mode. The measurement results with booster ramping in a full cycle,
before and after compensation, are shown in Fig. \ref{fig:3-7-FELImg_BefAft_AC}.
Without compensation, the average perturbation of the image center
is about 770 $\mu$m (average over measurements between $t=4.1$ s
and $t=6.6$ s) when the booster is parked at 1.2 GeV. This value
is significantly reduced to about $-3$ $\mu$m (average) with the
compensation, which is within the level of the rms noise (58 $\mu$m,
from data between $t=0\text{ to }2\ \text{s and }$$t=8\text{ to }10\text{ s}$
when the booster is operated at the lowest energy of 169 MeV). Along
the booster ramp-down process, a small but visible bump shows up in
the negative direction. The height of this peak is about 1/4 of the
uncompensated value with the booster operated at 1.2 GeV, which is
consistent with the previous BPM measurement result.

\noindent \begin{center}
\includegraphics[width=8cm]{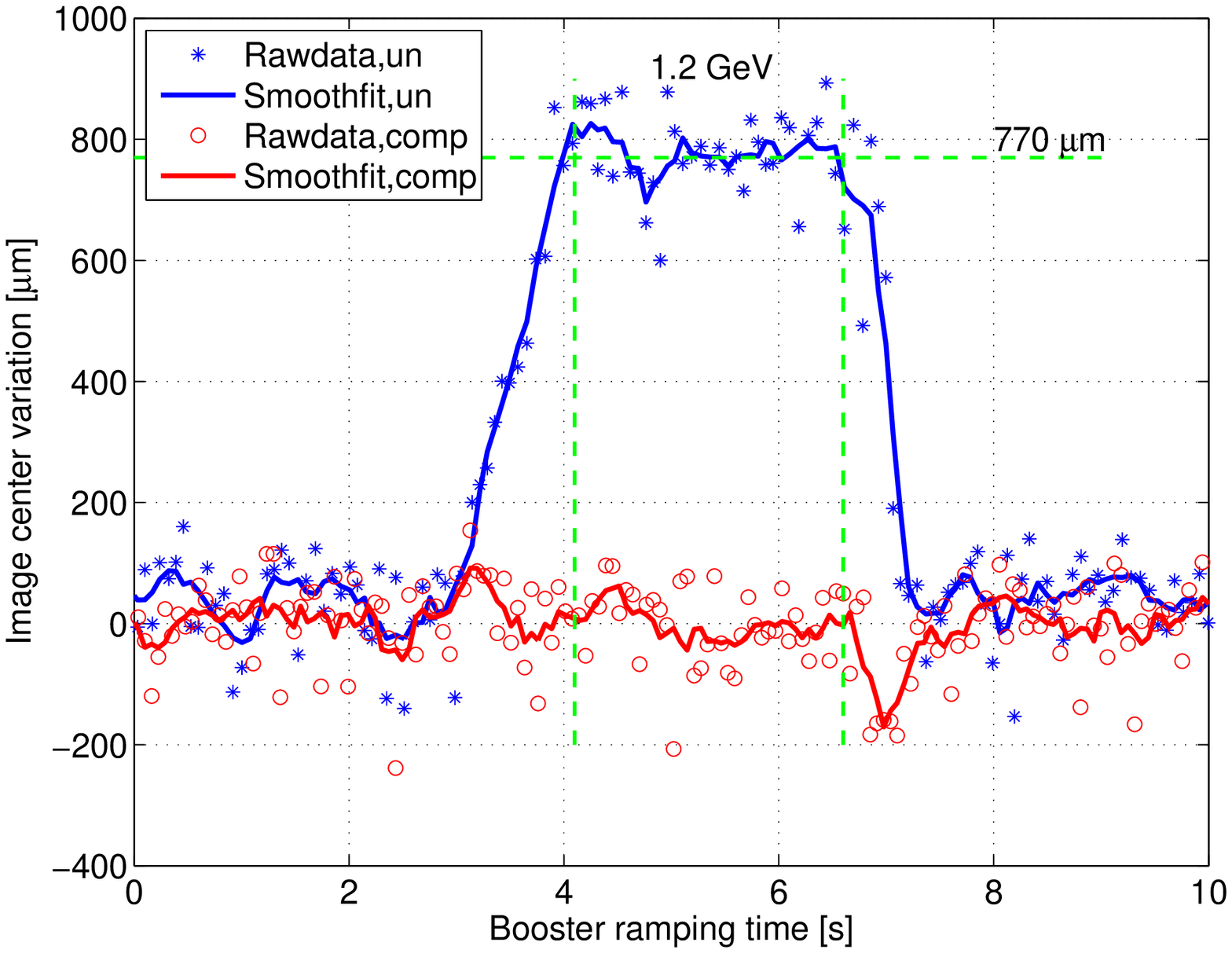}\figcaption{(color
online) Variations of undulator radiation image center measured using
a beam profile monitor during a full booster ramp cycle, the booster
is parked at 1.2 GeV for about 2.5 s as in Fig. \ref{fig:3-2-Rmsorb_BefAft_AC}.
Blue stars are the measured beam centers without compensation; red-circles
are those measured with compensation; blue-line and red-line are the
smoothed curves using a 5-point moving average. $E{}_{SR}=426$ MeV.\label{fig:3-7-FELImg_BefAft_AC}}
\par\end{center}

\section{{\normalsize{}Summary and discussion}}

The stability of the electron beam orbit is a critical requirement
for the HIGS operation at the Duke storage ring. To keep the aiming
of the gamma-ray beam stable, therefore to maintain the energy resolution
of the highly collimated gamma-ray beam, the e-beam orbit distortion
in the storage ring, especially at the Compton collision point, needs
to be minimized. In the development of the compensation scheme for
the booster leakage field, two effective correctors are chosen using
the beam based technique. Faster air coil correctors without steel
yokes are developed and installed, and their strengths for orbit correction
are determined by stepping the booster through a set of energies between
169 MeV and 1.2 GeV. The field compensation is implemented as a feedforward
in the realtime booster control system.

With this compensation scheme, the maximum perturbation of the horizontal
beam orbit and gamma-ray beam aiming is significantly decreased by
a factor of 3 with the booster operated in the ramping mode. In the
routine operation, the overall time-integrated distortion in the gamma-ray
beam aiming is expected to be decreased by a factor of 14 to 25. The projected
maximum horizontal gamma-ray beam center shift at the collimator is
reduced from 1677 $\mu$m to 447 $\mu$m (about 340 $\mu$m in the
vertical without the need for compensation) with booster ramping and
for the lowest energy operation of the storage ring (240 MeV). This
value is about 7\% of the smallest collimator diameter ($D=6$ mm),
well below our preliminary goal of 10\%. We plan to improve the orbit
compensation during the booster ramp-down process. Instead of using
energy-indexed corrector settings measured by stepping up the booster,
a new set of settings will be determined by stepping down the booster
energy.\\

\emph{We would like to thank the engineering and technical staff at
DFELL/TUNL for their support of this research work. This work was
supported by }National Natural Science Foundation of China (No. 11175180,
11475167)\emph{ and DOE Grant (No. DE-FG02-97ER41033). One of the
authors (Wei Li) also would like to thank the China Scholarship Council
(CSC) for supporting his research visit at Duke University.}

\end{multicols}

\vspace{-1mm}
 \centerline{\rule{80mm}{0.1pt}} \vspace{2mm}
 \begin{multicols}{2}

\end{multicols}

\clearpage{}

\end{CJK*}
\end{document}